\documentclass[twocolumn]{aastex6311}
\usepackage{threeparttable}
\usepackage{multirow}
\usepackage{times}
\usepackage{enumitem}
\usepackage{natbib}
\usepackage{url} 
\usepackage{amsmath,bm}
\usepackage{multirow}

\shorttitle{Stellar Parameters and Abundances for SMSS DR4 Stars}
\shortauthors{Yang Huang et al.}

\begin{document}

\title{Stellar Parameters for over Fifty Million stars from SMSS DR4 and Gaia DR3}

\author[0000-0001-8424-1079]{Yang Huang}
\affiliation{School of Astronomy and Space Science, University of Chinese Academy of Sciences, Beijing 100049, People's Republic of China}
\affiliation{CAS Key Lab of Optical Astronomy, National Astronomical Observatories, Chinese Academy of Sciences, Beijing 100012, People's Republic of China}

\author[0000-0003-4573-6233]{Timothy C. Beers}
\affiliation{Department of Physics and Astronomy and JINA Center for the Evolution of the Elements (JINA-CEE), University of Notre Dame, Notre Dame, IN 46556, USA}

\begin{abstract}
We present an updated catalog of stellar parameters, including effective temperature, luminosity classification, and metallicity, for over fifty million stars from the SkyMapper Southern Survey (SMSS) DR4 and \textit{Gaia} DR3. The accuracy of the derived parameters remains consistent with those achieved with SMSS DR2 using the same methods. Thanks to the advancements in SMSS DR4, photometric-metallicity estimates are now available for an unprecedented number of metal-poor stars. The catalog includes over 13 million metal-poor (MP; [Fe/H] $\leq -1$) stars, nearly three million very metal-poor (VMP; [Fe/H] $\leq -2.0$) stars, and approximately 120,000 extremely metal-poor (EMP; [Fe/H] $\leq -3.0$) stars -- representing an increase by a factor of 4–6 compared to SMSS DR2. This catalog, combined with other stellar parameters obtained through our efforts, will be made available at \url{https://nadc.china-vo.org/data/sports/} and \url{https://zenodo.org/records/15108911}.
\end{abstract}

\section{Introduction}
Obtaining a large sample of stars with accurate stellar parameters is essential for studying the formation and evolution of our Galaxy. To leverage the vast number of Galactic stars with photometric and astrometric measurements from \textit{Gaia}, \citet{PaperI} proposed deriving stellar parameters, particularly metallicity, from narrow- and medium-/wide-band photometric surveys. Using data from the SkyMapper Southern Survey (SMSS) DR2 \citep{SMSSDR2} and \textit{Gaia} EDR3 \citep{GEDR3}, they derived stellar parameters for over 20 million stars, four to five times more than those observed by LAMOST \citep{lamost, 2012RAA....12..723Z}, the largest spectroscopic survey of the Milky Way -- while achieving a metallicity precision comparable to that of LAMOST.  

Expanding on this approach, \citet{PaperII} derived stellar parameters for another 20 million stars in the Northern Hemisphere using data from SAGES DR1 \citep{SAGES} and \textit{Gaia} EDR3 \citep{GEDR3}. \citet{PaperIII} employed the Javalambre Photometric Local Universe Survey (J-PLUS) DR3 \citep{jplus} and \textit{Gaia} EDR3 \citep{GEDR3} to derive not only stellar parameters, but also several elemental-abundance ratios ([C/Fe], [Mg/Fe], and [$\alpha$/Fe]), for nearly five million stars. Similar efforts are presently underway for stars observed with the Southern Photometric Local Universe Survey (S-PLUS) DR4 \citep{2019MNRAS.489..241M}.   

Most recently, the SkyMapper Southern Survey (SMSS) released its fourth and final data release \citep[hereafter SMSS DR4;][]{SMSSDR4}. Compared to previous releases, SMSS DR4 features expanded sky coverage (26,000 square degrees compared to 21,000 square degrees in DR2) and a significantly larger number of unique objects (724 million targets versus 285 million in DR2), based on over 400,000 images collected between 2014 and 2021.  
The 10$\sigma$ depth in the $u-$ and $v-$ bands reaches 18.6 and 18.9 mag, respectively.  
Furthermore, photometric zero-points have been examined and corrected using \textit{Gaia} XP spectra \citep{2023A&A...674A...2D}, confirming the calibration issues previously identified by \citet{2021ApJ...907...68H} through application of the stellar 
color-regression method.

Given the significant progress made by SMSS DR4, in this Note, we apply the methods developed in \citet{PaperI} to SMSS DR4 and \textit{Gaia} DR3 in order to expand the number of stars with derived stellar parameters.

\begin{figure*}
\begin{center}
\includegraphics[scale=0.325,angle=0]{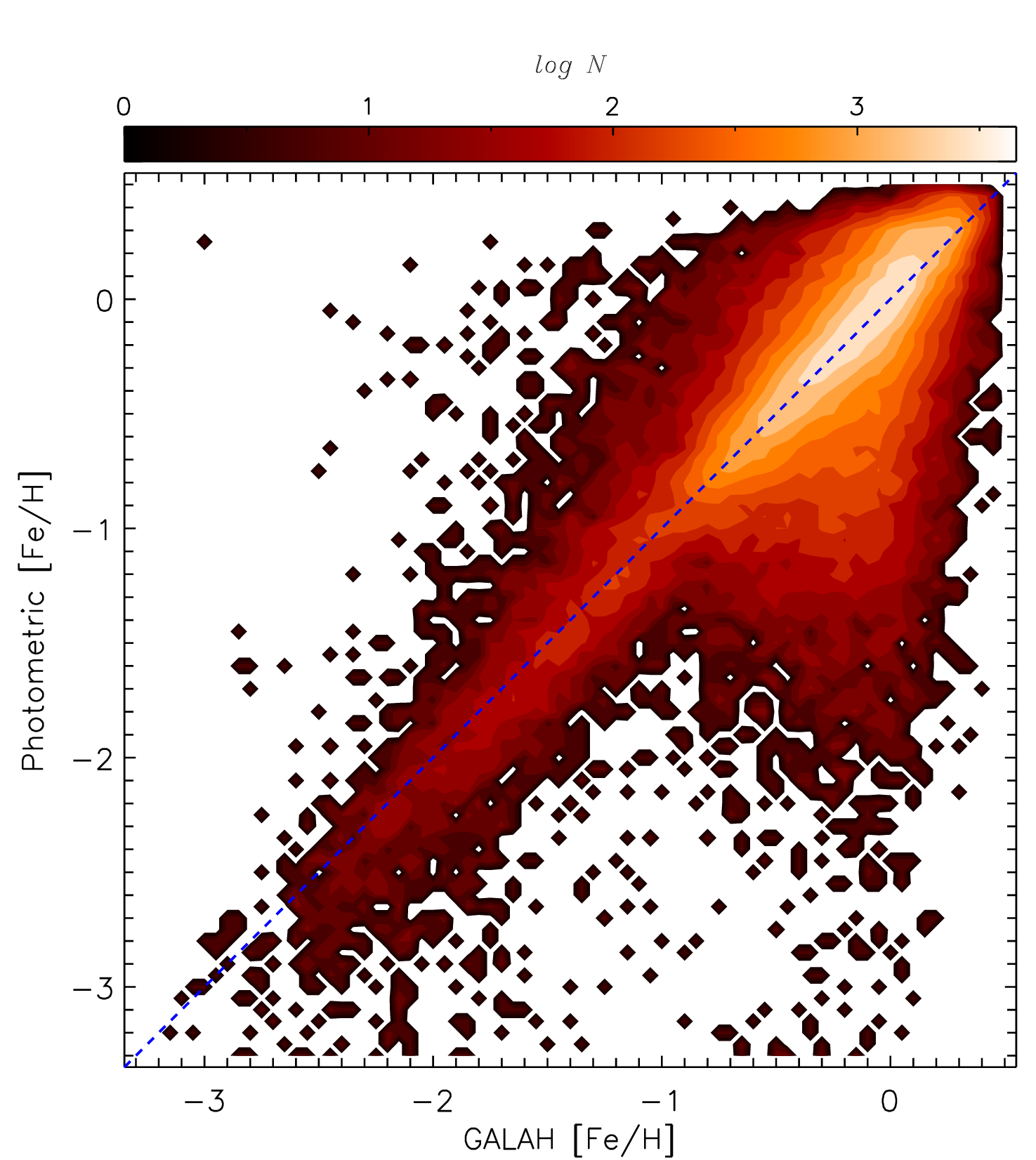}
\includegraphics[scale=0.325,angle=0]{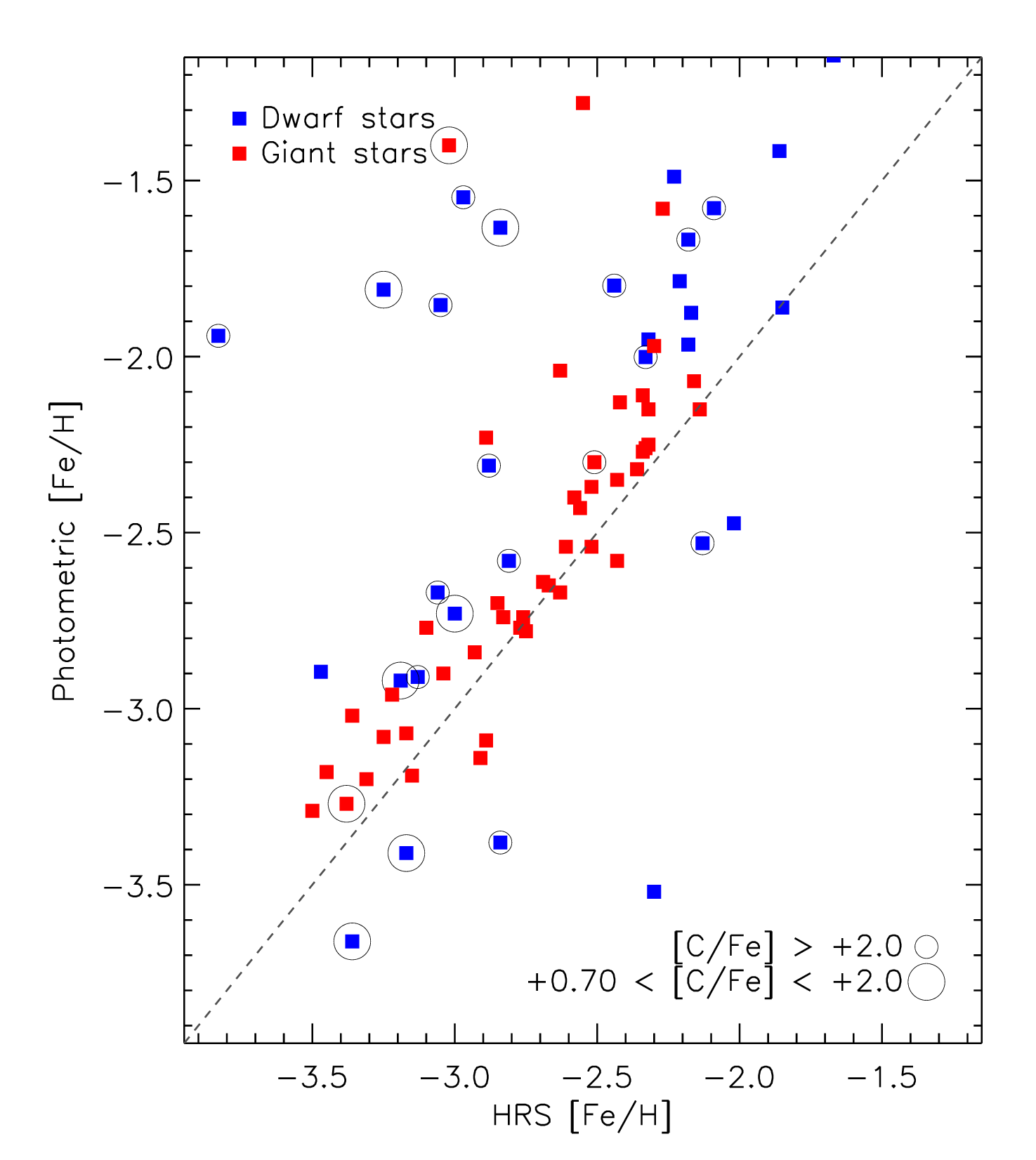}
\caption{\textbf{Left panel:} Comparison of photometric metallicity estimates with those from GALAH DR4 \citep{2024arXiv240919858B}. The color-coded contours represent stellar number density. The median offset and scatter in metallicity differences between GALAH DR4 and our photometric estimates are $-0.02$~dex and $0.12$~dex, respectively.  
\textbf{Right panel:} Comparison of photometric-metallicity estimates with high-resolution spectroscopic (HRS) metallicities from the GALAH DR4 database for dwarf (red squares) and giant (blue squares) stars. The small and large circles denote stars with carbon enhancements ($+0.7 < \text{[C/Fe]} \leq +2.0$) and extreme carbon enhancements ($\text{[C/Fe]} > +2.0$), respectively. The central dashed line represents the one-to-one correspondence. The median offset and scatter in metallicity differences between the HRS and photometric estimates are $-0.17$~dex and $0.26$~dex, respectively. When excluding carbon-enhanced metal-poor stars with $\text{[C/Fe]} > +0.7$, the median offset and scatter improve to $-0.10$~dex and $0.24$~dex, respectively.}
\end{center}
\end{figure*} 

\section{Data Selection}
We first select stars in SMSS DR4 with good photometric quality in the $u$- and $v$- bands using the following criteria: 1) {\tt class\_star $\geq 0.6$}, and 2) {\tt u/v\_flags $\leq 3$} and {\tt u/v\_ngood $\geq 1$}. This winnowing process results in a sample of over 85 million stars.  
We then cross-match these stars with \textit{Gaia} EDR3, and find that the overwhelming majority (99.2\%) have both \textit{Gaia} photometric and astrometric information.
For reddening correction, we adopted the 2-D map from \citet{SFD98} for stars at high Galactic latitudes ($|b| \geq 10^\circ$), and the 3-D map from \citet{2019ApJ...887...93G} for stars at low Galactic latitudes ($|b| < 10^\circ$).

\section{An updated catalog of stellar parameters}

Similar to \citet{PaperI}, we first train metallicity-dependent stellar loci using third-order 2-D polynomials for $G_{\rm BP} - G_{\rm RP}$ versus $u - G_{\rm BP}$ and $G_{\rm BP} - G_{\rm RP}$ versus $v - G_{\rm BP}$.  
The calibration labels are adopted from \citet{PaperIII}.  
Using these newly established stellar loci, we successfully derived photometric metallicities for nearly 39 million dwarf stars and over 13 million giant stars.  
Compared to metallicity estimates from GALAH DR4 \citep{2024arXiv240919858B}, the offset is negligible, and the scatter is only 0.12~dex (see the left panel of Fig.~1).  
Focusing on the very metal-poor regime ([Fe/H] $\leq -2.0$), our photometric metallicities exhibit excellent agreement with high-resolution spectroscopic measurements \citep{2022ApJ...931..147L}, with a moderate offset of approximately $-0.2$~dex and a small scatter of about 0.25~dex.  
We note that the metallicities of carbon-enhanced metal-poor (CEMP) stars tend to be over-estimated, due to contamination of the $u$- and $v$-band filters by molecular carbon bands, consistent with previous findings in \citet{PaperI, PaperII}.  
This issue can be effectively mitigated by incorporating carbon-sensitive medium-band filters, such as those available in J/S-PLUS \citep{PaperIII}.  

Based on the estimated photometric metallicities, we further derive photometric effective temperatures, distances for distant stars lacking reliable {\it Gaia} parallaxes, and stellar ages for stars with well-determined distances from {\it Gaia} parallax.  
Building on this dataset, along with previous works from our group \citep{PaperI, PaperII, PaperIII} and ongoing efforts, we have initiated a new project named SPORTS (Stellar Parameters fOR all The Stars; PIs: Yang Huang \& Timothy Beers). SPORTS represents a dedicated effort to determine the stellar parameters for all (or at least a large fraction of)  Milky Way stars using data from current and next-generation mega-surveys.  
The complete dataset of stellar parameters from SMSS DR4 will be accessible by August 30, 2025, at \url{https://nadc.china-vo.org/data/sports/} and will also be available for download from \url{https://zenodo.org/records/15108911}.
We note that, if this sample is used, please also cite \cite{PaperI}, as all the techniques applied are taken from that paper.

\vskip 2cm
 \section*{Acknowledgements} 
Y.H. acknowledges support by the National Natural Science Foundation of China grants No. 12422303 and the National Key R\&D Program of China (2023YFA1608303).
T.C.B. acknowledges partial support from grant PHY 14-30152; Physics Frontier Center/JINA Center for the Evolution of the Elements (JINA-CEE), and from OISE-1927130: The International Research Network for Nuclear Astrophysics (IReNA), awarded by the US National Science Foundation.

\vfill\eject
\bibliography{sppara_calib}{}
\bibliographystyle{aasjournal}
\end{document}